\documentclass{article}
\usepackage{arxiv}

\usepackage[utf8]{inputenc} 
\usepackage[T1]{fontenc}    
\usepackage{hyperref}       
\usepackage{url}            
\usepackage{booktabs}       
\usepackage{amsfonts}       
\usepackage{nicefrac}       
\usepackage{microtype}      
\usepackage{lipsum}

\usepackage[table]{xcolor}

\usepackage{float}
\usepackage{graphicx}

\title{VisMaker: a Question-Oriented Visualization Recommender System for Data Exploration}

\author{
  Raul~de~Araújo~Lima \\
  Department of Informatics, PUC-Rio\\
  Rua Marques de Sao Vicente 225 \\
  Rio de Janeiro, RJ, 22420-030, Brazil \\
  \texttt{rlima@inf.puc-rio.br} \\
   \And
  Simone~Diniz~Junqueira~Barbosa \\
  Department of Informatics, PUC-Rio\\
  Rua Marques de Sao Vicente 225 \\
  Rio de Janeiro, RJ, 22420-030, Brazil \\
  \texttt{simone@inf.puc-rio.br} \\
}

\usepackage{multirow}

\usepackage{soul}

\begin{document}
\maketitle

\begin{abstract}
The increasingly rapid growth of data production and the consequent need to explore data to obtain answers to the most varied questions have promoted the development of tools to facilitate the manipulation and construction of data visualizations. However, building useful data visualizations is not a trivial task: it may involve a large number of subtle decisions that require experience from their designer. In this paper, we present \textit{VisMaker}, a visualization recommender tool that uses a set of rules to present visualization recommendations organized and described through questions, in order to facilitate the understanding of the recommendations and assisting the visual exploration process. We carried out two studies comparing our tool with \textit{Voyager~2} and analyzed some aspects of the use of tools. We collected feedback from participants to identify the advantages and disadvantages of our recommendation approach. As a result, we gathered comments to help improve the development of tools in this domain.
\end{abstract}

\keywords{visualization recommendation \and visual data exploration \and visualization tool \and information visualization}

\section{Introduction}
The increasingly rapid growth of data production and the resulting need to explore data to obtain knowledge and answers to the most varied questions have promoted the development of tools to facilitate the data exploration process through the construction of data visualizations.

Since data visualizations aim \textit{``to aid our understanding of data by leveraging the human visual systems highly tuned ability to see patterns, spot trends, and identify outliers''}~\cite{heer2010tour}, they should enable readers to effectively explore datasets, communicating information more accurately and providing greater knowledge gain about the underlying data.

According to Pinker (1990), all data visualizations communicate a set of mathematical values through objects with visual dimensions (\textit{i.e.}, length, position, etc.) that correspond to the respective values~\cite{pinker1990theory}. They have a ``DNA'': a mapping between data and visual dimensions; and the most different visualization types can be constructed by varying these mappings~\cite{heer2010tour}. However, the process of constructing a data visualization requires a large number of decisions that involve the specification of which question to ask and what data to use, besides the selection of the most appropriate visual encodings to present the data~\cite{heer2010tour}. The selection of visual dimensions is, in many cases, the main barrier for the construction of useful data visualizations, especially by novices~\cite{grammel2010information}.

Many works have been developed in the Information Visualization (InfoVis) research area. Some of them have discussed about visualization effectiveness, proposing ranks of different visual encoding (sometimes called channels) when used to communicate some types of statistical variables~\cite{cleveland1984graphical, mackinlay1986automating, munzner2014visualization}. More recently, some works have presented tools for constructing or recommending data visualizations through rules based on the state-of-the-art of visualization effectiveness, helping users to explore datasets through the construction of visualizations~\cite{de2014recommender, gotz2009behavior, wongsuphasawat2016voyager, wongsuphasawat2017voyager}.
	

In this paper, we present VisMaker, a visualization recommender tool that recommends data visualizations grouped by questions that they answer, facilitating the user's understanding of the recommendations, also increasing their interest. We compared our tool with Voyager~2~\cite{wongsuphasawat2017voyager}, another tool in the same domain and with a similar purpose. This comparison was conducted in two different studies: (i)~the first one focused on question answering and (ii)~the second one focused on a data exploration scenario. The obtained results show that the VisMaker recommendation approach can be very useful for users who, through the questions presented by the tool,  find it easier to understand the recommended charts. VisMaker also assists in the process of investigating hypotheses about the data during the exploration process.

This paper is organized as follows: Section~\ref{sec:related-works} presents some related works that address visualization recommendations. The VisMaker tool is presented in Section~\ref{sec:vismaker}. Section~\ref{sec:evaluation} describes the comparison studies that we conducted to evaluate VisMaker. In Section~\ref{sec:discussion}, we discuss the obtained results. Finally, Section~\ref{sec:conclusion} presents our conclusions remarks and future works.

\section{Related Works}
\label{sec:related-works}

This paper presents a new tool to advance the state-of-the-art in visualization recommendation. In this field, many tools and works have been developed, starting at general-purpose tools for creating and manipulating spreadsheets, such as Microsoft Excel and Google Spreadsheets, which also enable the construction of data visualizations and are well known by many computer users.


One of the first tool to specify and automatically design effective graphical representations was the APT system~\cite{mackinlay1986automating}. It uses a set of primitive graphical languages and expressiveness and effectiveness criteria, and proposes a compositional algebra to enumerate possible visualizations. More recently, tools such as Tableau~\cite{stolte2002polaris}\footnote{\url{https://www.tableau.com/}}, have been developed to enable the construction of data visualizations more efficiently, where the users can specify the visualization by defining which visualization channels will be responsible for presenting which attribute. Tableau also provides recommendations for different visualization types, through the \textit{``Show Me''} interface~\cite{mackinlay2007show}.

Gotz and Wen (2009) developed a visualization recommender system guided by user behavior. The system consists of two phases: pattern detection and visualization recommendation~\cite{gotz2009behavior}. In the first phase, the system identifies user interaction patterns, and in second phase the system uses the detected patterns to infer the visual task that the user intends to perform on the data, recommending alternative visualizations. 

The \textit{VizDeck} tool~\cite{perry2013vizdeck} automatically recommends a set of visualizations based on the statistical properties of the data. Visualizations can be fetched by a search field through which the user can enter visualization types or attribute names by filtering the visualizations.

Sousa and Barbosa\cite{de2014recommender} developed the \textit{ViSC} tool, a recommender system to support chart construction for statistical data. ViSC uses an ontology that defines the task that the user can perform on the data, and maps it onto questions and visualizations that may be used to answer it. The system works as follows: first, the user selects which data will be visualized; the system presents a visualization deemed appropriate for the user and lists some questions related to the selected data and the current visualization, each one linked to a new visualization.

More recently,~\cite{wongsuphasawat2016voyager} presented \textit{Voyager}, a visualization recommender tool that provides a list of possible visualizations constructed from the variable selection, performed by the user, through the addition of other variables and the application of functions and data transformations. Voyager was compared to PoleStart\footnote{\url{https://vega.github.io/polestar/}}, which was previously constructed to generate visualizations based on the mapping between variables and visual dimensions. The authors also constructed a set of mappings that are used to define which visualization should present the set of selected variables. As an evolution of \textit{Voyager}, \cite{wongsuphasawat2017voyager} presented \textit{Voyager 2},\footnote{\url{https://vega.github.io/voyager/}} which, in addition to recommending visualizations based on user variables selection, allows users to specify visualizations by defining the mapping between variables and visual dimensions, blending manual and automatic chart specification.

Other works have used machine learning and optimization approaches to determine which visualization types are most appropriate and efficient to present data. \cite{dibia2018data2vis} formulated visualization projection as a translation problem between data specifications and visualization specifications. The work demonstrated the feasibility of this approach through the results obtained by the model and performed the integration of the model with a web tool that recommends a list of visualizations for a given user-specified dataset.

The DeepEye system~\cite{luo2018deepeye} is a tool for automatic building data visualization which addresses three issues: (i)~visualization recognition, (ii)~visualization ranking, and (iii)~visualization selection. The authors constructed several visualization datasets and used them to train some Machine Learning models to perform the visualization recognition task and to rank visualizations. The VizML system, in turn, used a dataset of 120,000 tables and related visualizations. The authors developed a Machine Learning model to classify and recommend visualizations according to an attribute combination~\cite{vizml}.

As we describe in the next section, VisMaker is a new tool that, similar to Voyager~2, uses a combination of manual and automatic visualization specifications. The significant difference of VisMaker is that the presentation and recommendation of visualizations are organized by questions that aim to facilitate the understanding of the recommendations, helping the user to make and test hypotheses about a given dataset during data exploration, thus reducing their \textit{``visual mapping barrier''}~\cite{grammel2010information}.


\section{The VisMaker Tool}
\label{sec:vismaker}

In this section, we present the VisMaker tool. We introduce its user interface and the mappings between data, questions and visualization types, used to drive the visualization recommendation process.




\subsection{Technologies Used by VisMaker}

We developed VisMaker as a Web application using frameworks like Vue.js\footnote{\url{https://vuejs.org/}} and BootstrapVue\footnote{\url{https://bootstrap-vue.js.org/}}. We also use Vega-lite\footnote{\url{https://vega.github.io/vega-lite/}}~\cite{satyanarayan2016vega}, a specification language for constructing and embedding data visualizations.

Vega-lite was developed to be a high-level specification language to facilitate the visualization construction process through the low-level grammar Vega, presented in~\cite{satyanarayan2015reactive}. According to~\cite{wongsuphasawat2016voyager}, Vega-lite \textit{``consists of a set of mappings between visual encoding channels and (potentially transformed) data variables''}. Vega-lite specifications are defined as JSON (JavaScript Object Notation) objects, which are compiled into Vega specifications.




\subsection{User Interface}

\begin{figure}[htb]
	\centering
	\includegraphics[width=\textwidth]{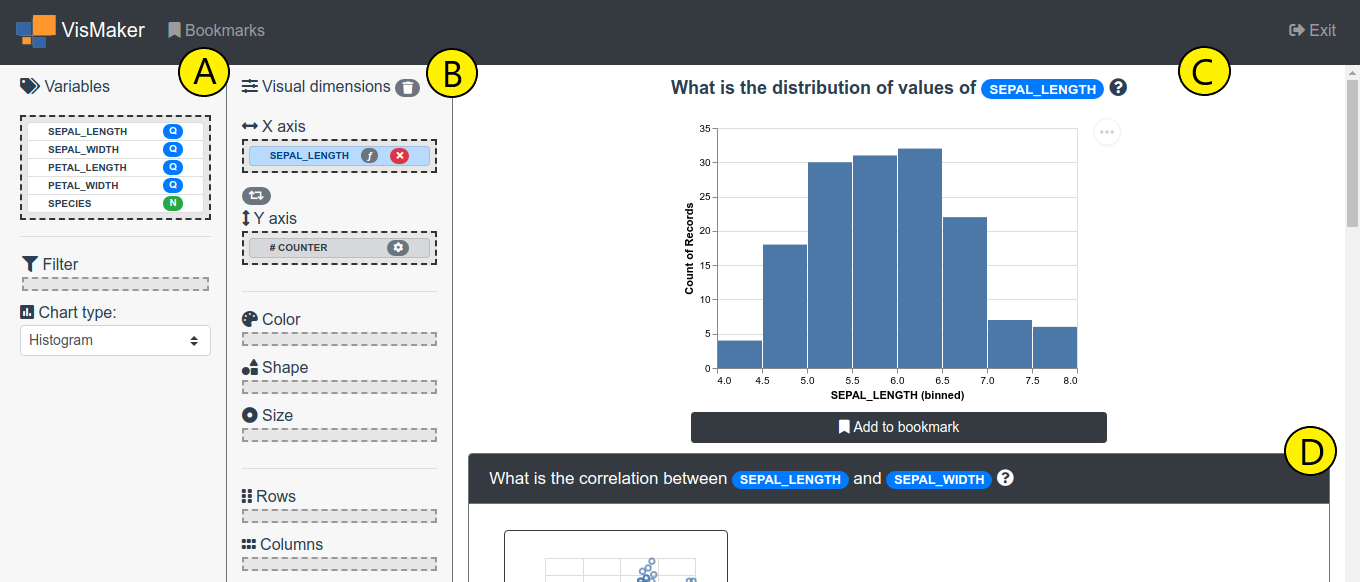}
	\caption{VisMaker main interface}
	\label{fig:vismaker_panels_1}
\end{figure}

Figure~\ref{fig:vismaker_panels_1} presents the VisMaker user interface after having loaded the Iris dataset and added the \texttt{sepal\_length} variable to the X-axis. In the figure, panel~(A) presents all the dataset variables and their types. Panel~(B) shows all the supported visual dimensions, represented as dropdown widgets. To define the visualization that will be presented in panel~(C), users drag the variables of interest from panel~(A) and drop them to the respective field in panel~(B). Panel~(D) presents a list of related questions and visualizations constructed by adding new variables to the current variable selection. 

After loading a dataset, panels (C) and (D) present, respectively, an empty chart and a list of questions and charts created by selection a single variable. In addition, only the X and Y axes in panel~(B) are available to receive mapped variables, since the other visual dimensions require first mapping at least one variable to the X or Y axis. In the following, we describe each panel.

\subsubsection{Variables Panel}

VisMaker currently considers only four variable types: quantitative~(Q), nominal~(N), ordinal~(O) and temporal~(T). We select a unique color that visually distinguishes the presentation of each variable type: blue for quantitative, green for nominal, and yellow for temporal variables. Through the variables panel, users can change the variable types by clicking on the colored button with the letter that identifies the variable type.

\subsubsection{Visual Dimensions Panel}

VisMaker provides a set of seven visual dimensions (X axis, Y axis, color, shape, size, rows, and columns), which represent some visualization channels and can be used to specify the main visualization through the Vega-Lite\footnote{\url{https://vega.github.io/}} grammar~\cite{satyanarayan2016vega}.

\subsubsection{Questions Panel} 

\begin{figure}
  \includegraphics[width=0.9\textwidth]{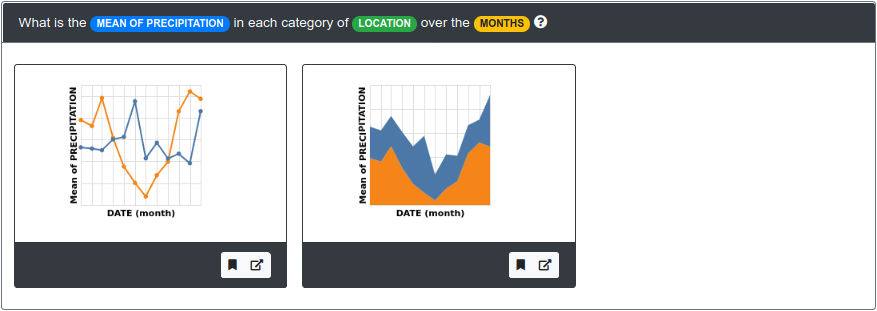}
  \caption{An example of a VisMaker question panel}
  \label{fig:vismaker_teaser}
\end{figure}

The questions panel presents a list of cards, illustrated in Figure~\ref{fig:vismaker_teaser}. The list contains a question and a set of visualizations that can be used to answer the question. Each visualization is presented in a card that has a bookmark button followed by a button that can be used to transfer the corresponding chart into the main visualization panel. In the Questions panel, we also use the color set for each variable type, highlighting the variable names whenever they appear in a question to facilitate their identification.

\subsection{Mappings between Data, Questions, and Visualizations}

The Tableau \textit{`Show Me'} interface~\cite{mackinlay2007show} used a mapping between variable types to automatically identify the visualization type. In turn, the Voyager tool proposes a mapping between variable types and the Vega-lite mark types. Vega-lite considers quantitative (Q), nominal (N), ordinal (O), and temporal (T) data. In~\cite{srinivasan2019augmenting}, some of the data analytic tasks identified by \cite{amar2005low} have been mapped onto a set of variable types to understand which visualizations and data facts should be presented when the user defines a combination of attributes. 


VisMaker has two recommender systems that uses a set of rules constructed based on those mappings. The first one consists of determining the most appropriate chart according to the mapping between variables and visualization encodings that the user has defined. It includes the automatic addition of aggregate summaries, which can anticipate some user actions, directly generating some visualizations that require the step of adding a `counter' attribute in other tools, like Voyager~2. For this recommender system, we constructed a mapping that defines which chart mark should be used according to variable type combinations for the X and Y axes. Table~\ref{tab:mapping-1} presents the mappings used in VisMaker.

\begin{table}[ht]
\centering
\caption{Mappings between variable types and chart types used in VisMaker}
\label{tab:mapping-1}
\begin{tabular}{@{}ll@{}}
\toprule
\textbf{Variabele types} & \textbf{Visualization types}      \\ \midrule
Q                          & Area chart; Histogram             \\
N                          & Bar chart                         \\
O                          & Line chart; Bar chart; Area chart \\
T                          & Line chart; Bar chart; Area chart \\
Q x Q                      & Scatter plot                      \\
Q x N                      & Box plot; Strip plot              \\
Q x O                      & Line chart; Bar chart; Area chart \\
Q x T                      & Line chart; Bar chart; Area chart \\
N x N                      & Heatmap                           \\
N x O                      & Heatmap                           \\
N x T                      & Heatmap                           \\
O x O                      & Heatmap                           \\
O x T                      & Heatmap                           \\
T x T                      & Heatmap                           \\ \bottomrule
\end{tabular}
\end{table}

The second recommender system is used to support users in data exploration processes. It is based on the \textit{Compass} recommendation engine used by Voyager~\cite{wongsuphasawat2016voyager}. This system feeds the question panel by adding a new variable from the unselected variables set. It generates recommendations through a set of predefined mappings between variable types, questions, and chart types. Table~\ref{tab:mapping-2} presents some of the most commonly used mappings we have defined. In VisMaker, the variable names are highlighted according to the color that identifies their type, as shown in Figure~\ref{fig:vismaker_teaser}, which presents an example of the second recommended question for Q x N x T types.

\begin{table*}[ht]
\centering
\caption{Mappings between data, questions and visualizations used in VisMaker}
\label{tab:mapping-2}
\resizebox{\textwidth}{!}{
\begin{tabular}{@{}lll@{}}
\toprule
\textbf{Variable types} & \textbf{Question} & \textbf{Chart types} \\ \midrule
Q x Q & 1. What is the correlation between \underline{\texttt{var0}} and \underline{\texttt{var1}}? & Scatterplot \\
\multirow{2}{*}{Q x N} & 1. What is the is the distribution of values of \underline{\texttt{var0}} in each category of \underline{\texttt{var1}}? & Boxplot; Strip plot; Area chart \\
 & 2. What is the average of \underline{\texttt{var0}} in each category of \underline{\texttt{var1}}? & Bar chart \\
\multirow{2}{*}{Q x T} & 1. What is the \underline{\texttt{MEAN OF var0}} over the \underline{\texttt{YEARS}}? & Line chart; Bar chart; Area chart \\
 & 2. What is the \underline{\texttt{MEAN OF var0}} over the \underline{\texttt{MONTHS}}? & Line chart; Bar chart; Area chart \\
N x N & 1. What is the number of co-occurrences between each category of \underline{\texttt{var0}} and \underline{\texttt{var1}}? & Stacked bar chart; Heatmap \\
\multirow{2}{*}{N x T} & 1. What is the number of occurrences of each category of \underline{\texttt{var0}} over the \underline{\texttt{YEARS}}? & Line chart; Stacked area chart; Heatmap \\
 & 2. What is the number of occurrences of each category of \underline{\texttt{var0}} over the \underline{\texttt{MONTHS}}? & Line chart; Stacked area chart; Heatmap \\ 
T x T & 1. What is number of co-occurrences of \underline{\texttt{var0}} and \underline{\texttt{var1}}? & Stacked bar chart; Heatmap \\
Q x Q x Q & 1. What is the correlation between \underline{\texttt{var0}}, \underline{\texttt{var1}} and \underline{\texttt{var2}}? & Scatterplot + color; Scatterplot + size \\
Q x Q x N & 1. What is the correlation between \underline{\texttt{var0}}, \underline{\texttt{var1}} grouped by \underline{\texttt{var2}} categories? & Scatterplot + color; Scatterplot + shape \\
\multirow{2}{*}{Q x Q x T} & 1. What is the correlation between \underline{\texttt{MEAN OF var0}}, \underline{\texttt{MEAN OF var1}} over the \underline{\texttt{YEARS}}? & Line chart + size; Line chart + color \\
 & 2. What is the correlation between \underline{\texttt{MEAN OF var0}}, \underline{\texttt{MEAN OF var1}} over the \underline{\texttt{MONTHS}}? & Line chart + size; Line chart + color \\
Q x N x N & 1. What is the \underline{\texttt{MEAN OF var0}} in each combination of \underline{\texttt{var1}} and \underline{\texttt{var2}}? & Heatmap \\
\multirow{2}{*}{Q x N x T} & 1. What is the \underline{\texttt{MEAN OF var0}} in each category of \underline{\texttt{var1}} over the \underline{\texttt{YEARS}}? & Line chart; Stacked area chart \\
 & 2. What is the \underline{\texttt{MEAN OF var0}} in each category of \underline{\texttt{var1}} over the \underline{\texttt{MONTHS}}? & Line chart; Stacked area chart \\
Q x T x T & 1. What is the \underline{\texttt{MEAN OF var0}} in each combination of \underline{\texttt{var1}} and \underline{\texttt{var2}}? & Line chart; Heatmap \\ \bottomrule
\end{tabular}
}
\end{table*}

\section{Evaluation}
\label{sec:evaluation}

We compared VisMaker and Voyager~2~\cite{wongsuphasawat2017voyager} through two different studies: the first study evaluated the tools when they are used in a question-answering scenario; and the second study, designed from a limitation observed in the first,  evaluated the tools when they are used in a data exploration scenario. We conducted the studies with 24 volunteer participants, 16 in the first study, and 8 in the second. Participants received two tasks, one to complete using each tool. They used the tools to construct visualizations to answer questions about data, in the case of the first study; and to explore and gain knowledge about the data, in the case of the second study.

For conducting these studies, we built an agenda through which participants could schedule a time for the session. All experiment sessions were conducted at the university. Participants used a Lenovo\textsuperscript{\textregistered} notebook with Intel\textsuperscript{\textregistered} Core\textsuperscript{\texttrademark} i5 processor and 8GB of RAM. 

We obtained two datasets that we used in the two studies. For the question-answering study (Section~\ref{sec:study1}), we prepared for each dataset a list of questions that participants should answer through data visualizations. The tasks are presented in Section~\ref{sec:tasks}. While in the first study participants had to answer questions and the tasks are composed of a dataset and a set of questions, in the second study, participants should freely build useful visualizations that make it possible to make interesting discoveries about the data.

We divided the participants into four distinct groups, varying the tool-task pair and the order in which each tool would be used. At the beginning of each session, we asked each participant to read and sign an informed consent form and, having accepted it, to fill out a participant characterization form, answering some questions that we considered necessary to define their profile. Then, participants set out to solve the tasks using a specific tool, according to the order and the tool-task pair defined for their group. After completing each task, participants answered a 7-point Likert scale questionnaire to evaluate the perceived ease of use and utility based on the Technology Acceptance Model (TAM)~\cite{davis1989perceived}.

As the focus of this study was not on usability, such as the discoverability of the tools, we introduced participants to the use of tools through video tutorials of approximately 10 minutes, before they started using them. We conducted a pilot study to get a sense of how long each session would take to complete, and we verified that it would take approximately 1 hour and 30 minutes.

\subsection{Tasks}
\label{sec:tasks}

We used the same datasets to compose the tasks of both studies. They are described below:

\subsubsection{The GRADUATE PROGRAMS task} 
This task consists of answering a set of 8 questions (4 manually constructed and 4 mainly through the tool recommendations) about a dataset of 4,993 records of graduate programs located in different cities of the state of Rio de Janeiro over the years. This dataset has seven variables: one of it is temporal (\texttt{YEAR}), three are nominal (\texttt{CITY}, \texttt{JURIDICAL STATUS} and \texttt{FIELD}), and three are quantitative (\texttt{QNT MASTERS}, \texttt{QNT DOCTORAL} and \texttt{QNT POSTDOCTORAL}). 



\subsubsection{The WEATHER task} 
The WEATHER task consists of answering also a set of 8 questions, but about a dataset of 2,922 weather records in the cities of New York and Seattle. This dataset has seven variables, one of which is temporal (\texttt{DATE}), two are nominal (\texttt{LOCATION} and \texttt{WEATHER}), and four are quantitative (\texttt{WIND}, \texttt{PRECIPITATION}, \texttt{TEMP\_MAX} and \texttt{TEMP\_MIN}).

\subsection{Study 1: Question Answering}
\label{sec:study1}

In this first study, participants were given tasks about a specific dataset and a set of questions to be answered through data visualizations. We did not specify a time limit for achieving each task. 

\subsubsection{Findings}

\begin{figure}
	\includegraphics[width=1\linewidth]{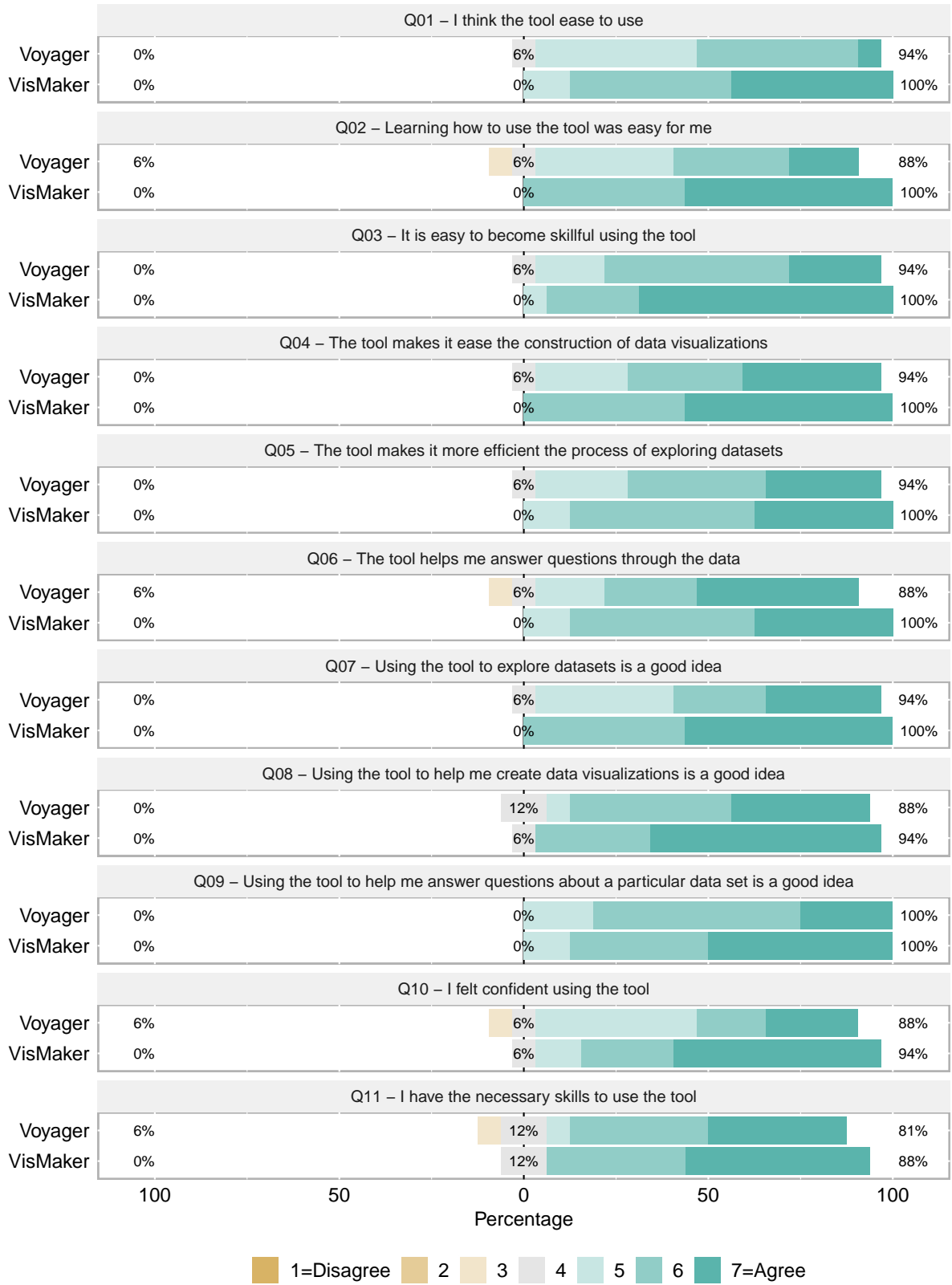}
	\caption{Summary of responses in Study 1}
	\label{fig:questions_1}
\end{figure}

Through the questionnaires answered by the participants after using each of the tools, we were able to gathered their opinions about some aspects of the tools. Figure~\ref{fig:questions_1} shows the distribution of the participants' answers to the 11 questions present in the questionnaires for each tool.

As the responses were provided in an ordinal scale, we used the Mann-Whitney~\cite{wilcoxon1992individual} test to verify the statistical significance of the answers for each tool. The results showed that, although VisMaker received greater or equal acceptance in most questions, the differences between the tools were only statistically significant ($p-value \geq 0.05$) for the answers to questions Q01, Q02, Q03, Q07, and Q10. Table~\ref{tab:statistics-1} presents the corresponding \textit{p}-values, as well as the medians of answers of each tool in each question.

\begin{table}[ht]
\centering
\caption{Results of Mann-Whitney statistical test for Study 1}
\label{tab:statistics-1}
\begin{tabular}{@{}lrrr@{}}
\toprule
\textbf{Question} & \textbf{VisMaker median} & \textbf{Voyager median} & \textbf{\textit{p}-value} \\ \midrule
Q01 & 6 & 5.5 & \textbf{0.005*} \\
Q02 & 7 & 5.5 & \textbf{0.002*} \\
Q03 & 7 & 6 & \textbf{0.013*} \\
Q04 & 7 & 6 & 0.092 \\
Q05 & 6 & 6 & 0.363 \\
Q06 & 6 & 6 & 0.702 \\
Q07 & 7 & 6 & \textbf{0.023*} \\
Q08 & 7 & 6 & 0.145 \\
Q09 & 6.5 & 6 & 0.195 \\
Q10 & 7 & 5 & \textbf{0.038*} \\
Q11 & 6.5 & 6 & 0.382 \\ \bottomrule
\end{tabular}
\end{table}

Regarding task correctness, we found four cases: (i)~those who built correct visualizations and responded correctly in the form; (ii)~those who built correct visualizations but provided an incorrect answer in the form; (iii)~those who built incorrect visualizations but answered correctly in the form; and (iv)~those who built incorrect visualizations and provided an incorrect answer in the form.

Tables~\ref{tab:correctness_visualizations} and \ref{tab:correctness_answers} show the correctness of the participants' visualizations and responses in the forms, respectively. The tables highlight the correct use of the VisMaker tool in gray, and Voyager in blue. Errors are pointed out in red.

\begin{table}[ht]
\centering
\caption{Correctness of the constructed visualizations}
\resizebox{\linewidth}{!}{
\begin{tabular}{|c|c|c|c|c|c|c|c|c|c|c|c|c|c|c|c|c|c|c|c|c|c|c|c|}
\hline
\multicolumn{6}{|c|}{\textbf{Group 1}} & \multicolumn{6}{c|}{\textbf{Group 2}} & \multicolumn{6}{c|}{\textbf{Group 3}} & \multicolumn{6}{c|}{\textbf{Group 4}} \\ \hline
\multicolumn{2}{|c|}{\textbf{Task}} & \textbf{P01} & \textbf{P02} & \textbf{P03} & \textbf{P04} & \multicolumn{2}{c|}{\textbf{Task}} & \textbf{P05} & \textbf{P06} & \textbf{P07} & \textbf{P08} & \multicolumn{2}{c|}{\textbf{Task}} & \textbf{P09} & \textbf{P10} & \textbf{P11} & \textbf{P12} & \multicolumn{2}{c|}{\textbf{Task}} & \textbf{P13} & \textbf{P14} & \textbf{P15} & \textbf{P16} \\ \hline
 & T1 & \cellcolor[HTML]{EFEFEF} & \cellcolor[HTML]{EFEFEF} & \cellcolor[HTML]{EFEFEF} & \cellcolor[HTML]{EFEFEF} &  & T1 & \cellcolor[HTML]{EFEFEF} & \cellcolor[HTML]{EFEFEF} & \cellcolor[HTML]{EFEFEF} & \cellcolor[HTML]{FD6864} &  & T1 & \cellcolor[HTML]{6FB3FF} & \cellcolor[HTML]{6FB3FF} & \cellcolor[HTML]{6FB3FF} & \cellcolor[HTML]{6FB3FF} &  & T1 & \cellcolor[HTML]{6FB3FF} & \cellcolor[HTML]{6FB3FF} & \cellcolor[HTML]{6FB3FF} & \cellcolor[HTML]{6FB3FF} \\ \cline{2-6} \cline{8-12} \cline{14-18} \cline{20-24} 
 & T2 & \cellcolor[HTML]{EFEFEF} & \cellcolor[HTML]{EFEFEF} & \cellcolor[HTML]{EFEFEF} & \cellcolor[HTML]{EFEFEF} &  & T2 & \cellcolor[HTML]{EFEFEF} & \cellcolor[HTML]{EFEFEF} & \cellcolor[HTML]{EFEFEF} & \cellcolor[HTML]{EFEFEF} &  & T2 & \cellcolor[HTML]{6FB3FF} & \cellcolor[HTML]{6FB3FF} & \cellcolor[HTML]{6FB3FF} & \cellcolor[HTML]{6FB3FF} &  & T2 & \cellcolor[HTML]{6FB3FF} & \cellcolor[HTML]{6FB3FF} & \cellcolor[HTML]{6FB3FF} & \cellcolor[HTML]{6FB3FF} \\ \cline{2-6} \cline{8-12} \cline{14-18} \cline{20-24} 
 & T3 & \cellcolor[HTML]{EFEFEF} & \cellcolor[HTML]{EFEFEF} & \cellcolor[HTML]{EFEFEF} & \cellcolor[HTML]{EFEFEF} &  & T3 & \cellcolor[HTML]{FD6864} & \cellcolor[HTML]{EFEFEF} & \cellcolor[HTML]{EFEFEF} & \cellcolor[HTML]{EFEFEF} &  & T3 & \cellcolor[HTML]{6FB3FF} & \cellcolor[HTML]{6FB3FF} & \cellcolor[HTML]{6FB3FF} & \cellcolor[HTML]{6FB3FF} &  & T3 & \cellcolor[HTML]{FD6864} & \cellcolor[HTML]{FD6864} & \cellcolor[HTML]{6FB3FF} & \cellcolor[HTML]{6FB3FF} \\ \cline{2-6} \cline{8-12} \cline{14-18} \cline{20-24} 
 & T4 & \cellcolor[HTML]{EFEFEF} & \cellcolor[HTML]{EFEFEF} & \cellcolor[HTML]{EFEFEF} & \cellcolor[HTML]{EFEFEF} &  & T4 & \cellcolor[HTML]{EFEFEF} & \cellcolor[HTML]{EFEFEF} & \cellcolor[HTML]{EFEFEF} & \cellcolor[HTML]{EFEFEF} &  & T4 & \cellcolor[HTML]{6FB3FF} & \cellcolor[HTML]{6FB3FF} & \cellcolor[HTML]{6FB3FF} & \cellcolor[HTML]{6FB3FF} &  & T4 & \cellcolor[HTML]{6FB3FF} & \cellcolor[HTML]{6FB3FF} & \cellcolor[HTML]{6FB3FF} & \cellcolor[HTML]{6FB3FF} \\ \cline{2-6} \cline{8-12} \cline{14-18} \cline{20-24} 
 & T5 & \cellcolor[HTML]{EFEFEF} & \cellcolor[HTML]{EFEFEF} & \cellcolor[HTML]{EFEFEF} & \cellcolor[HTML]{EFEFEF} &  & T5 & \cellcolor[HTML]{EFEFEF} & \cellcolor[HTML]{EFEFEF} & \cellcolor[HTML]{EFEFEF} & \cellcolor[HTML]{EFEFEF} &  & T5 & \cellcolor[HTML]{6FB3FF} & \cellcolor[HTML]{6FB3FF} & \cellcolor[HTML]{6FB3FF} & \cellcolor[HTML]{6FB3FF} &  & T5 & \cellcolor[HTML]{6FB3FF} & \cellcolor[HTML]{6FB3FF} & \cellcolor[HTML]{6FB3FF} & \cellcolor[HTML]{6FB3FF} \\ \cline{2-6} \cline{8-12} \cline{14-18} \cline{20-24} 
 & T6 & \cellcolor[HTML]{EFEFEF} & \cellcolor[HTML]{EFEFEF} & \cellcolor[HTML]{EFEFEF} & \cellcolor[HTML]{EFEFEF} &  & T6 & \cellcolor[HTML]{EFEFEF} & \cellcolor[HTML]{EFEFEF} & \cellcolor[HTML]{EFEFEF} & \cellcolor[HTML]{EFEFEF} &  & T6 & \cellcolor[HTML]{6FB3FF} & \cellcolor[HTML]{6FB3FF} & \cellcolor[HTML]{6FB3FF} & \cellcolor[HTML]{6FB3FF} &  & T6 & \cellcolor[HTML]{6FB3FF} & \cellcolor[HTML]{6FB3FF} & \cellcolor[HTML]{6FB3FF} & \cellcolor[HTML]{6FB3FF} \\ \cline{2-6} \cline{8-12} \cline{14-18} \cline{20-24} 
 & T7 & \cellcolor[HTML]{EFEFEF} & \cellcolor[HTML]{EFEFEF} & \cellcolor[HTML]{EFEFEF} & \cellcolor[HTML]{EFEFEF} &  & T7 & \cellcolor[HTML]{EFEFEF} & \cellcolor[HTML]{EFEFEF} & \cellcolor[HTML]{EFEFEF} & \cellcolor[HTML]{EFEFEF} &  & T7 & \cellcolor[HTML]{6FB3FF} & \cellcolor[HTML]{6FB3FF} & \cellcolor[HTML]{6FB3FF} & \cellcolor[HTML]{6FB3FF} &  & T7 & \cellcolor[HTML]{6FB3FF} & \cellcolor[HTML]{6FB3FF} & \cellcolor[HTML]{6FB3FF} & \cellcolor[HTML]{6FB3FF} \\ \cline{2-6} \cline{8-12} \cline{14-18} \cline{20-24} 
\multirow{-8}{*}{\rotatebox[origin=c]{90}{\textbf{GRAD. PROG.}}} & T8 & \cellcolor[HTML]{EFEFEF} & \cellcolor[HTML]{EFEFEF} & \cellcolor[HTML]{EFEFEF} & \cellcolor[HTML]{EFEFEF} & \multirow{-8}{*}{\rotatebox[origin=c]{90}{\textbf{WEATHER}}} & T8 & \cellcolor[HTML]{EFEFEF} & \cellcolor[HTML]{EFEFEF} & \cellcolor[HTML]{EFEFEF} & \cellcolor[HTML]{EFEFEF} & \multirow{-8}{*}{\rotatebox[origin=c]{90}{\textbf{GRAD. PROG.}}} & T8 & \cellcolor[HTML]{6FB3FF} & \cellcolor[HTML]{6FB3FF} & \cellcolor[HTML]{6FB3FF} & \cellcolor[HTML]{6FB3FF} & \multirow{-8}{*}{\rotatebox[origin=c]{90}{\textbf{WEATHER}}} & T8 & \cellcolor[HTML]{6FB3FF} & \cellcolor[HTML]{6FB3FF} & \cellcolor[HTML]{6FB3FF} & \cellcolor[HTML]{6FB3FF} \\ \hline
 & T1 & \cellcolor[HTML]{6FB3FF} & \cellcolor[HTML]{6FB3FF} & \cellcolor[HTML]{6FB3FF} & \cellcolor[HTML]{6FB3FF} &  & T1 & \cellcolor[HTML]{6FB3FF} & \cellcolor[HTML]{6FB3FF} & \cellcolor[HTML]{FD6864} & \cellcolor[HTML]{6FB3FF} &  & T1 & \cellcolor[HTML]{EFEFEF} & \cellcolor[HTML]{EFEFEF} & \cellcolor[HTML]{EFEFEF} & \cellcolor[HTML]{EFEFEF} &  & T1 & \cellcolor[HTML]{EFEFEF} & \cellcolor[HTML]{EFEFEF} & \cellcolor[HTML]{EFEFEF} & \cellcolor[HTML]{EFEFEF} \\ \cline{2-6} \cline{8-12} \cline{14-18} \cline{20-24} 
 & T2 & \cellcolor[HTML]{6FB3FF} & \cellcolor[HTML]{6FB3FF} & \cellcolor[HTML]{6FB3FF} & \cellcolor[HTML]{6FB3FF} &  & T2 & \cellcolor[HTML]{6FB3FF} & \cellcolor[HTML]{6FB3FF} & \cellcolor[HTML]{6FB3FF} & \cellcolor[HTML]{6FB3FF} &  & T2 & \cellcolor[HTML]{EFEFEF} & \cellcolor[HTML]{EFEFEF} & \cellcolor[HTML]{EFEFEF} & \cellcolor[HTML]{EFEFEF} &  & T2 & \cellcolor[HTML]{EFEFEF} & \cellcolor[HTML]{EFEFEF} & \cellcolor[HTML]{EFEFEF} & \cellcolor[HTML]{EFEFEF} \\ \cline{2-6} \cline{8-12} \cline{14-18} \cline{20-24} 
 & T3 & \cellcolor[HTML]{6FB3FF} & \cellcolor[HTML]{6FB3FF} & \cellcolor[HTML]{6FB3FF} & \cellcolor[HTML]{6FB3FF} &  & T3 & \cellcolor[HTML]{6FB3FF} & \cellcolor[HTML]{6FB3FF} & \cellcolor[HTML]{FD6864} & \cellcolor[HTML]{6FB3FF} &  & T3 & \cellcolor[HTML]{EFEFEF} & \cellcolor[HTML]{EFEFEF} & \cellcolor[HTML]{EFEFEF} & \cellcolor[HTML]{EFEFEF} &  & T3 & \cellcolor[HTML]{FD6864} & \cellcolor[HTML]{EFEFEF} & \cellcolor[HTML]{EFEFEF} & \cellcolor[HTML]{EFEFEF} \\ \cline{2-6} \cline{8-12} \cline{14-18} \cline{20-24} 
 & T4 & \cellcolor[HTML]{6FB3FF} & \cellcolor[HTML]{6FB3FF} & \cellcolor[HTML]{6FB3FF} & \cellcolor[HTML]{FD6864} &  & T4 & \cellcolor[HTML]{6FB3FF} & \cellcolor[HTML]{6FB3FF} & \cellcolor[HTML]{6FB3FF} & \cellcolor[HTML]{FD6864} &  & T4 & \cellcolor[HTML]{EFEFEF} & \cellcolor[HTML]{EFEFEF} & \cellcolor[HTML]{EFEFEF} & \cellcolor[HTML]{EFEFEF} &  & T4 & \cellcolor[HTML]{EFEFEF} & \cellcolor[HTML]{EFEFEF} & \cellcolor[HTML]{EFEFEF} & \cellcolor[HTML]{EFEFEF} \\ \cline{2-6} \cline{8-12} \cline{14-18} \cline{20-24} 
 & T5 & \cellcolor[HTML]{6FB3FF} & \cellcolor[HTML]{6FB3FF} & \cellcolor[HTML]{6FB3FF} & \cellcolor[HTML]{6FB3FF} &  & T5 & \cellcolor[HTML]{6FB3FF} & \cellcolor[HTML]{6FB3FF} & \cellcolor[HTML]{6FB3FF} & \cellcolor[HTML]{6FB3FF} &  & T5 & \cellcolor[HTML]{EFEFEF} & \cellcolor[HTML]{EFEFEF} & \cellcolor[HTML]{EFEFEF} & \cellcolor[HTML]{EFEFEF} &  & T5 & \cellcolor[HTML]{FD6864} & \cellcolor[HTML]{EFEFEF} & \cellcolor[HTML]{EFEFEF} & \cellcolor[HTML]{EFEFEF} \\ \cline{2-6} \cline{8-12} \cline{14-18} \cline{20-24} 
 & T6 & \cellcolor[HTML]{6FB3FF} & \cellcolor[HTML]{6FB3FF} & \cellcolor[HTML]{6FB3FF} & \cellcolor[HTML]{6FB3FF} &  & T6 & \cellcolor[HTML]{6FB3FF} & \cellcolor[HTML]{6FB3FF} & \cellcolor[HTML]{6FB3FF} & \cellcolor[HTML]{6FB3FF} &  & T6 & \cellcolor[HTML]{EFEFEF} & \cellcolor[HTML]{EFEFEF} & \cellcolor[HTML]{EFEFEF} & \cellcolor[HTML]{EFEFEF} &  & T6 & \cellcolor[HTML]{EFEFEF} & \cellcolor[HTML]{EFEFEF} & \cellcolor[HTML]{EFEFEF} & \cellcolor[HTML]{EFEFEF} \\ \cline{2-6} \cline{8-12} \cline{14-18} \cline{20-24} 
 & T7 & \cellcolor[HTML]{6FB3FF} & \cellcolor[HTML]{6FB3FF} & \cellcolor[HTML]{6FB3FF} & \cellcolor[HTML]{6FB3FF} &  & T7 & \cellcolor[HTML]{6FB3FF} & \cellcolor[HTML]{6FB3FF} & \cellcolor[HTML]{FD6864} & \cellcolor[HTML]{6FB3FF} &  & T7 & \cellcolor[HTML]{EFEFEF} & \cellcolor[HTML]{EFEFEF} & \cellcolor[HTML]{EFEFEF} & \cellcolor[HTML]{EFEFEF} &  & T7 & \cellcolor[HTML]{EFEFEF} & \cellcolor[HTML]{EFEFEF} & \cellcolor[HTML]{EFEFEF} & \cellcolor[HTML]{EFEFEF} \\ \cline{2-6} \cline{8-12} \cline{14-18} \cline{20-24} 
\multirow{-8}{*}{\rotatebox[origin=c]{90}{\textbf{WEATHER}}} & T8 & \cellcolor[HTML]{6FB3FF} & \cellcolor[HTML]{6FB3FF} & \cellcolor[HTML]{6FB3FF} & \cellcolor[HTML]{6FB3FF} & \multirow{-8}{*}{\rotatebox[origin=c]{90}{\textbf{GRAD. PROG.}}} & T8 & \cellcolor[HTML]{6FB3FF} & \cellcolor[HTML]{6FB3FF} & \cellcolor[HTML]{6FB3FF} & \cellcolor[HTML]{FD6864} & \multirow{-8}{*}{\rotatebox[origin=c]{90}{\textbf{WEATHER}}} & T8 & \cellcolor[HTML]{EFEFEF} & \cellcolor[HTML]{EFEFEF} & \cellcolor[HTML]{EFEFEF} & \cellcolor[HTML]{EFEFEF} & \multirow{-8}{*}{\rotatebox[origin=c]{90}{\textbf{GRAD. PROG.}}} & T8 & \cellcolor[HTML]{EFEFEF} & \cellcolor[HTML]{EFEFEF} & \cellcolor[HTML]{EFEFEF} & \cellcolor[HTML]{EFEFEF} \\ \hline
\end{tabular}
}
\label{tab:correctness_visualizations}
\end{table}

\begin{table}[ht]
\centering
\caption{Correctness of the participants' answers to the task questions}
\resizebox{\linewidth}{!}{
\begin{tabular}{|c|c|c|c|c|c|c|c|c|c|c|c|c|c|c|c|c|c|c|c|c|c|c|c|}
\hline
\multicolumn{6}{|c|}{\textbf{Group 1}} & \multicolumn{6}{c|}{\textbf{Group 2}} & \multicolumn{6}{c|}{\textbf{Group 3}} & \multicolumn{6}{c|}{\textbf{Group 4}} \\ \hline
\multicolumn{2}{|c|}{\textbf{Task}} & \textbf{P01} & \textbf{P02} & \textbf{P03} & \textbf{P04} & \multicolumn{2}{c|}{\textbf{Task}} & \textbf{P05} & \textbf{P06} & \textbf{P07} & \textbf{P08} & \multicolumn{2}{c|}{\textbf{Task}} & \textbf{P09} & \textbf{P10} & \textbf{P11} & \textbf{P12} & \multicolumn{2}{c|}{\textbf{Task}} & \textbf{P13} & \textbf{P14} & \textbf{P15} & \textbf{P16} \\ \hline
 & T1 & \cellcolor[HTML]{FD6864} & \cellcolor[HTML]{EFEFEF} & \cellcolor[HTML]{EFEFEF} & \cellcolor[HTML]{EFEFEF} &  & T1 & \cellcolor[HTML]{EFEFEF} & \cellcolor[HTML]{EFEFEF} & \cellcolor[HTML]{EFEFEF} & \cellcolor[HTML]{FD6864} &  & T1 & \cellcolor[HTML]{6FB3FF} & \cellcolor[HTML]{6FB3FF} & \cellcolor[HTML]{6FB3FF} & \cellcolor[HTML]{6FB3FF} &  & T1 & \cellcolor[HTML]{6FB3FF} & \cellcolor[HTML]{6FB3FF} & \cellcolor[HTML]{6FB3FF} & \cellcolor[HTML]{6FB3FF} \\ \cline{2-6} \cline{8-12} \cline{14-18} \cline{20-24} 
 & T2 & \cellcolor[HTML]{EFEFEF} & \cellcolor[HTML]{EFEFEF} & \cellcolor[HTML]{EFEFEF} & \cellcolor[HTML]{EFEFEF} &  & T2 & \cellcolor[HTML]{FD6864} & \cellcolor[HTML]{EFEFEF} & \cellcolor[HTML]{EFEFEF} & \cellcolor[HTML]{EFEFEF} &  & T2 & \cellcolor[HTML]{6FB3FF} & \cellcolor[HTML]{6FB3FF} & \cellcolor[HTML]{6FB3FF} & \cellcolor[HTML]{6FB3FF} &  & T2 & \cellcolor[HTML]{FD6864} & \cellcolor[HTML]{6FB3FF} & \cellcolor[HTML]{FD6864} & \cellcolor[HTML]{FD6864} \\ \cline{2-6} \cline{8-12} \cline{14-18} \cline{20-24} 
 & T3 & \cellcolor[HTML]{EFEFEF} & \cellcolor[HTML]{EFEFEF} & \cellcolor[HTML]{EFEFEF} & \cellcolor[HTML]{EFEFEF} &  & T3 & \cellcolor[HTML]{FD6864} & \cellcolor[HTML]{EFEFEF} & \cellcolor[HTML]{EFEFEF} & \cellcolor[HTML]{EFEFEF} &  & T3 & \cellcolor[HTML]{6FB3FF} & \cellcolor[HTML]{6FB3FF} & \cellcolor[HTML]{6FB3FF} & \cellcolor[HTML]{6FB3FF} &  & T3 & \cellcolor[HTML]{FD6864} & \cellcolor[HTML]{FD6864} & \cellcolor[HTML]{6FB3FF} & \cellcolor[HTML]{6FB3FF} \\ \cline{2-6} \cline{8-12} \cline{14-18} \cline{20-24} 
 & T4 & \cellcolor[HTML]{EFEFEF} & \cellcolor[HTML]{EFEFEF} & \cellcolor[HTML]{EFEFEF} & \cellcolor[HTML]{EFEFEF} &  & T4 & \cellcolor[HTML]{EFEFEF} & \cellcolor[HTML]{EFEFEF} & \cellcolor[HTML]{EFEFEF} & \cellcolor[HTML]{EFEFEF} &  & T4 & \cellcolor[HTML]{6FB3FF} & \cellcolor[HTML]{6FB3FF} & \cellcolor[HTML]{6FB3FF} & \cellcolor[HTML]{6FB3FF} &  & T4 & \cellcolor[HTML]{6FB3FF} & \cellcolor[HTML]{6FB3FF} & \cellcolor[HTML]{6FB3FF} & \cellcolor[HTML]{6FB3FF} \\ \cline{2-6} \cline{8-12} \cline{14-18} \cline{20-24} 
 & T5 & \cellcolor[HTML]{EFEFEF} & \cellcolor[HTML]{EFEFEF} & \cellcolor[HTML]{FD6864} & \cellcolor[HTML]{EFEFEF} &  & T5 & \cellcolor[HTML]{EFEFEF} & \cellcolor[HTML]{EFEFEF} & \cellcolor[HTML]{EFEFEF} & \cellcolor[HTML]{EFEFEF} &  & T5 & \cellcolor[HTML]{6FB3FF} & \cellcolor[HTML]{6FB3FF} & \cellcolor[HTML]{6FB3FF} & \cellcolor[HTML]{6FB3FF} &  & T5 & \cellcolor[HTML]{6FB3FF} & \cellcolor[HTML]{6FB3FF} & \cellcolor[HTML]{6FB3FF} & \cellcolor[HTML]{6FB3FF} \\ \cline{2-6} \cline{8-12} \cline{14-18} \cline{20-24} 
 & T6 & \cellcolor[HTML]{EFEFEF} & \cellcolor[HTML]{EFEFEF} & \cellcolor[HTML]{EFEFEF} & \cellcolor[HTML]{EFEFEF} &  & T6 & \cellcolor[HTML]{EFEFEF} & \cellcolor[HTML]{EFEFEF} & \cellcolor[HTML]{EFEFEF} & \cellcolor[HTML]{EFEFEF} &  & T6 & \cellcolor[HTML]{6FB3FF} & \cellcolor[HTML]{6FB3FF} & \cellcolor[HTML]{6FB3FF} & \cellcolor[HTML]{6FB3FF} &  & T6 & \cellcolor[HTML]{6FB3FF} & \cellcolor[HTML]{6FB3FF} & \cellcolor[HTML]{6FB3FF} & \cellcolor[HTML]{6FB3FF} \\ \cline{2-6} \cline{8-12} \cline{14-18} \cline{20-24} 
 & T7 & \cellcolor[HTML]{EFEFEF} & \cellcolor[HTML]{EFEFEF} & \cellcolor[HTML]{EFEFEF} & \cellcolor[HTML]{EFEFEF} &  & T7 & \cellcolor[HTML]{EFEFEF} & \cellcolor[HTML]{EFEFEF} & \cellcolor[HTML]{EFEFEF} & \cellcolor[HTML]{EFEFEF} &  & T7 & \cellcolor[HTML]{6FB3FF} & \cellcolor[HTML]{6FB3FF} & \cellcolor[HTML]{6FB3FF} & \cellcolor[HTML]{6FB3FF} &  & T7 & \cellcolor[HTML]{6FB3FF} & \cellcolor[HTML]{6FB3FF} & \cellcolor[HTML]{6FB3FF} & \cellcolor[HTML]{6FB3FF} \\ \cline{2-6} \cline{8-12} \cline{14-18} \cline{20-24} 
\multirow{-8}{*}{\rotatebox[origin=c]{90}{\textbf{GRAD. PROG.}}} & T8 & \cellcolor[HTML]{EFEFEF} & \cellcolor[HTML]{EFEFEF} & \cellcolor[HTML]{EFEFEF} & \cellcolor[HTML]{EFEFEF} & \multirow{-8}{*}{\rotatebox[origin=c]{90}{\textbf{WEATHER}}} & T8 & \cellcolor[HTML]{EFEFEF} & \cellcolor[HTML]{EFEFEF} & \cellcolor[HTML]{EFEFEF} & \cellcolor[HTML]{EFEFEF} & \multirow{-8}{*}{\rotatebox[origin=c]{90}{\textbf{GRAD. PROG.}}} & T8 & \cellcolor[HTML]{6FB3FF} & \cellcolor[HTML]{6FB3FF} & \cellcolor[HTML]{6FB3FF} & \cellcolor[HTML]{6FB3FF} & \multirow{-8}{*}{\rotatebox[origin=c]{90}{\textbf{WEATHER}}} & T8 & \cellcolor[HTML]{6FB3FF} & \cellcolor[HTML]{6FB3FF} & \cellcolor[HTML]{6FB3FF} & \cellcolor[HTML]{6FB3FF} \\ \hline
 & T1 & \cellcolor[HTML]{6FB3FF} & \cellcolor[HTML]{6FB3FF} & \cellcolor[HTML]{6FB3FF} & \cellcolor[HTML]{6FB3FF} &  & T1 & \cellcolor[HTML]{6FB3FF} & \cellcolor[HTML]{6FB3FF} & \cellcolor[HTML]{FD6864} & \cellcolor[HTML]{6FB3FF} &  & T1 & \cellcolor[HTML]{EFEFEF} & \cellcolor[HTML]{EFEFEF} & \cellcolor[HTML]{EFEFEF} & \cellcolor[HTML]{EFEFEF} &  & T1 & \cellcolor[HTML]{EFEFEF} & \cellcolor[HTML]{FD6864} & \cellcolor[HTML]{EFEFEF} & \cellcolor[HTML]{FD6864} \\ \cline{2-6} \cline{8-12} \cline{14-18} \cline{20-24} 
 & T2 & \cellcolor[HTML]{FD6864} & \cellcolor[HTML]{6FB3FF} & \cellcolor[HTML]{FD6864} & \cellcolor[HTML]{FD6864} &  & T2 & \cellcolor[HTML]{FD6864} & \cellcolor[HTML]{6FB3FF} & \cellcolor[HTML]{6FB3FF} & \cellcolor[HTML]{6FB3FF} &  & T2 & \cellcolor[HTML]{EFEFEF} & \cellcolor[HTML]{EFEFEF} & \cellcolor[HTML]{EFEFEF} & \cellcolor[HTML]{EFEFEF} &  & T2 & \cellcolor[HTML]{EFEFEF} & \cellcolor[HTML]{EFEFEF} & \cellcolor[HTML]{EFEFEF} & \cellcolor[HTML]{EFEFEF} \\ \cline{2-6} \cline{8-12} \cline{14-18} \cline{20-24} 
 & T3 & \cellcolor[HTML]{6FB3FF} & \cellcolor[HTML]{6FB3FF} & \cellcolor[HTML]{6FB3FF} & \cellcolor[HTML]{6FB3FF} &  & T3 & \cellcolor[HTML]{6FB3FF} & \cellcolor[HTML]{6FB3FF} & \cellcolor[HTML]{6FB3FF} & \cellcolor[HTML]{6FB3FF} &  & T3 & \cellcolor[HTML]{EFEFEF} & \cellcolor[HTML]{EFEFEF} & \cellcolor[HTML]{EFEFEF} & \cellcolor[HTML]{EFEFEF} &  & T3 & \cellcolor[HTML]{EFEFEF} & \cellcolor[HTML]{EFEFEF} & \cellcolor[HTML]{EFEFEF} & \cellcolor[HTML]{EFEFEF} \\ \cline{2-6} \cline{8-12} \cline{14-18} \cline{20-24} 
 & T4 & \cellcolor[HTML]{6FB3FF} & \cellcolor[HTML]{6FB3FF} & \cellcolor[HTML]{FD6864} & \cellcolor[HTML]{FD6864} &  & T4 & \cellcolor[HTML]{FD6864} & \cellcolor[HTML]{6FB3FF} & \cellcolor[HTML]{6FB3FF} & \cellcolor[HTML]{FD6864} &  & T4 & \cellcolor[HTML]{EFEFEF} & \cellcolor[HTML]{EFEFEF} & \cellcolor[HTML]{EFEFEF} & \cellcolor[HTML]{EFEFEF} &  & T4 & \cellcolor[HTML]{EFEFEF} & \cellcolor[HTML]{EFEFEF} & \cellcolor[HTML]{EFEFEF} & \cellcolor[HTML]{EFEFEF} \\ \cline{2-6} \cline{8-12} \cline{14-18} \cline{20-24} 
 & T5 & \cellcolor[HTML]{6FB3FF} & \cellcolor[HTML]{6FB3FF} & \cellcolor[HTML]{6FB3FF} & \cellcolor[HTML]{6FB3FF} &  & T5 & \cellcolor[HTML]{6FB3FF} & \cellcolor[HTML]{6FB3FF} & \cellcolor[HTML]{FD6864} & \cellcolor[HTML]{6FB3FF} &  & T5 & \cellcolor[HTML]{EFEFEF} & \cellcolor[HTML]{EFEFEF} & \cellcolor[HTML]{EFEFEF} & \cellcolor[HTML]{EFEFEF} &  & T5 & \cellcolor[HTML]{FD6864} & \cellcolor[HTML]{EFEFEF} & \cellcolor[HTML]{EFEFEF} & \cellcolor[HTML]{EFEFEF} \\ \cline{2-6} \cline{8-12} \cline{14-18} \cline{20-24} 
 & T6 & \cellcolor[HTML]{6FB3FF} & \cellcolor[HTML]{6FB3FF} & \cellcolor[HTML]{6FB3FF} & \cellcolor[HTML]{6FB3FF} &  & T6 & \cellcolor[HTML]{6FB3FF} & \cellcolor[HTML]{6FB3FF} & \cellcolor[HTML]{FD6864} & \cellcolor[HTML]{6FB3FF} &  & T6 & \cellcolor[HTML]{EFEFEF} & \cellcolor[HTML]{EFEFEF} & \cellcolor[HTML]{EFEFEF} & \cellcolor[HTML]{EFEFEF} &  & T6 & \cellcolor[HTML]{FD6864} & \cellcolor[HTML]{EFEFEF} & \cellcolor[HTML]{EFEFEF} & \cellcolor[HTML]{EFEFEF} \\ \cline{2-6} \cline{8-12} \cline{14-18} \cline{20-24} 
 & T7 & \cellcolor[HTML]{6FB3FF} & \cellcolor[HTML]{6FB3FF} & \cellcolor[HTML]{6FB3FF} & \cellcolor[HTML]{6FB3FF} &  & T7 & \cellcolor[HTML]{6FB3FF} & \cellcolor[HTML]{6FB3FF} & \cellcolor[HTML]{6FB3FF} & \cellcolor[HTML]{6FB3FF} &  & T7 & \cellcolor[HTML]{EFEFEF} & \cellcolor[HTML]{EFEFEF} & \cellcolor[HTML]{EFEFEF} & \cellcolor[HTML]{EFEFEF} &  & T7 & \cellcolor[HTML]{EFEFEF} & \cellcolor[HTML]{EFEFEF} & \cellcolor[HTML]{EFEFEF} & \cellcolor[HTML]{EFEFEF} \\ \cline{2-6} \cline{8-12} \cline{14-18} \cline{20-24} 
\multirow{-8}{*}{\rotatebox[origin=c]{90}{\textbf{WEATHER}}} & T8 & \cellcolor[HTML]{6FB3FF} & \cellcolor[HTML]{6FB3FF} & \cellcolor[HTML]{6FB3FF} & \cellcolor[HTML]{6FB3FF} & \multirow{-8}{*}{\rotatebox[origin=c]{90}{\textbf{GRAD. PROG.}}} & T8 & \cellcolor[HTML]{6FB3FF} & \cellcolor[HTML]{6FB3FF} & \cellcolor[HTML]{6FB3FF} & \cellcolor[HTML]{FD6864} & \multirow{-8}{*}{\rotatebox[origin=c]{90}{\textbf{WEATHER}}} & T8 & \cellcolor[HTML]{EFEFEF} & \cellcolor[HTML]{EFEFEF} & \cellcolor[HTML]{EFEFEF} & \cellcolor[HTML]{EFEFEF} & \multirow{-8}{*}{\rotatebox[origin=c]{90}{\textbf{GRAD. PROG.}}} & T8 & \cellcolor[HTML]{EFEFEF} & \cellcolor[HTML]{EFEFEF} & \cellcolor[HTML]{EFEFEF} & \cellcolor[HTML]{EFEFEF} \\ \hline
\end{tabular} }
\label{tab:correctness_answers}
\end{table}

We were able to group all the errors found in a few categories. The errors that occurred with the Voyager tool relate mainly to problems with the scale of the generated charts and to the use of variables. Most of the problems with the charts scale occurred during the resolution of the second question (T2) of the WEATHER task. Many of the participants in groups 1 and 4, who used Voyager to solve the WEATHER task, built tiny scale visualizations, which made it impossible to compare values, inducing them to provide an incorrect answer.


In VisMaker, some errors resulted from a bug in the tool. In these cases, participants constructed a visualization correctly, mapping variables and visual dimensions, but obtained ``distorted'' visualizations. These errors occurred mainly in the first question (T1) of the GRADUATE PROGRAMS task by three participants (P01, P14, and P16). In these cases, the participants used the DATE variable to filter only the year 2008 required by the question, and the tool presented as incorrect value.

\subsubsection{Participants' feedback}

We list below some feedback given by the participants for the main question we asked in the post-session interview in Study: \textbf{\textit{Considering the recommendation approaches, which tool helped you most to complete the task?}}
\begin{itemize}
    \item[P01:] \textit{``I found the two tools very similar in terms of recommendations. Still, for me, \textbf{VisMaker was more intuitive} because it presented a written question, and it helped me identify what I was looking for.''}
    
    \item[P02:] \textit{``I felt \textbf{more confident with Voyager, perhaps because it was the second tool I used}. Although the charts were worse to analyze, I thought that building them was more comfortable with Voyager. As the questionnaire of the task was guiding me, I always looked for the variables.''}
    
    \item[P03:] \textit{``\textbf{Voyager helped me more}. I did well on it because there were so many recommendations, and \textbf{despite having a more cluttered interface} with more information on the screen, I just needed to find one of the visualizations that were being recommended to answer the question.''}
    
    \item[P04:] \textit{``\textbf{I hardly used the recommendations}. I found it pretty much the same.''}
    
    \item[]P05:] \textit{``The way \textbf{VisMaker} presents recommendations is a summary of what visualization is, and it \textbf{seemed to me somewhat friendlier}, gave it a certain consistency that was the point I missed most in Voyager.''}
    
    \item[P06:] \textit{``\textbf{VisMaker was easier than Voyager}. Having the question helps, but I looked straight at the charts... When I had some difficulty, I used the questions to confirm the meaning of the charts.''}
    
    \item[P07:] \textit{``\textbf{VisMaker helped me more} because, instead of looking for information inside the chart, I was looking for questions that resembled the question I wanted to answer''}.
    
    \item[P08:] \textit{``\textbf{The two were similar, but I preferred VisMaker} because I found its way to present a little better. [...] \textbf{The fact that VisMaker presents the questions helps}, so I liked it more.''}
    
    \item[P09:] \textit{``\textbf{VisMaker helped because of the recommendation questions}.''}
    
    \item[P10:] \textit{``In general, \textbf{I liked Voyager, but the VisMaker recommendation approach was better for me because it presented the questions}. [...] I think \textbf{VisMaker is better when I need to use many variables} because the more variables I have to use, the more I have to keep asking the question I want to answer in my mind. \textbf{How VisMaker presents the questions it makes it easier}.''}
    
    \item[P11:] \textit{``\textbf{VisMaker helped me more because it already gave a better explanation} of why it was using certain variables in the form of questions.''}
    
    \item[P12:] \textit{``I think \textbf{VisMaker helped me more because it made the chart easier to understand, it was easier}.''}
    
    \item[P13:] \textit{``In VisMaker, sometimes, I had to clear the mappings and start over to see if the recommendations helped me get where I wanted. I think \textbf{Voyager bothered me less}.''}
    
    \item[P14:] \textit{``I found the recommendations of both tools quite similar, but the\textbf{ presentation of VisMaker is much better because the variables are highlighted by color}.''}
    
    \item[P15:] \textit{``In \textbf{VisMaker, having a question associated with the visualization increased my confidence} about what the visualization answered.''}
    
    \item[P16:] \textit{``Voyager shows you everything, and you don't have a categorization or a specialization of what is being recommended... \textbf{VisMaker was more specific in recommendations, and they were more accurate}.''}
\end{itemize}






\subsection{Study 2: Data Exploration}

In the second study, which aimed to evaluate the use of VisMaker in a data exploration scenario, when users do not have specific questions to answer, we used the same datasets as the first study and the same session structure.

The tasks that make up this study are based on the first study tasks described in Section~\ref{sec:tasks}, one dataset. Instead of answering a set of pre-established questions, participants had the objective of ``comprehensively exploring the data'' using the tools, generating useful visualizations. As the task was open ended, we set a 30-minute time limit.

\subsubsection{Findings}

\begin{figure}[H]
	\centering
	\includegraphics[width=\linewidth]{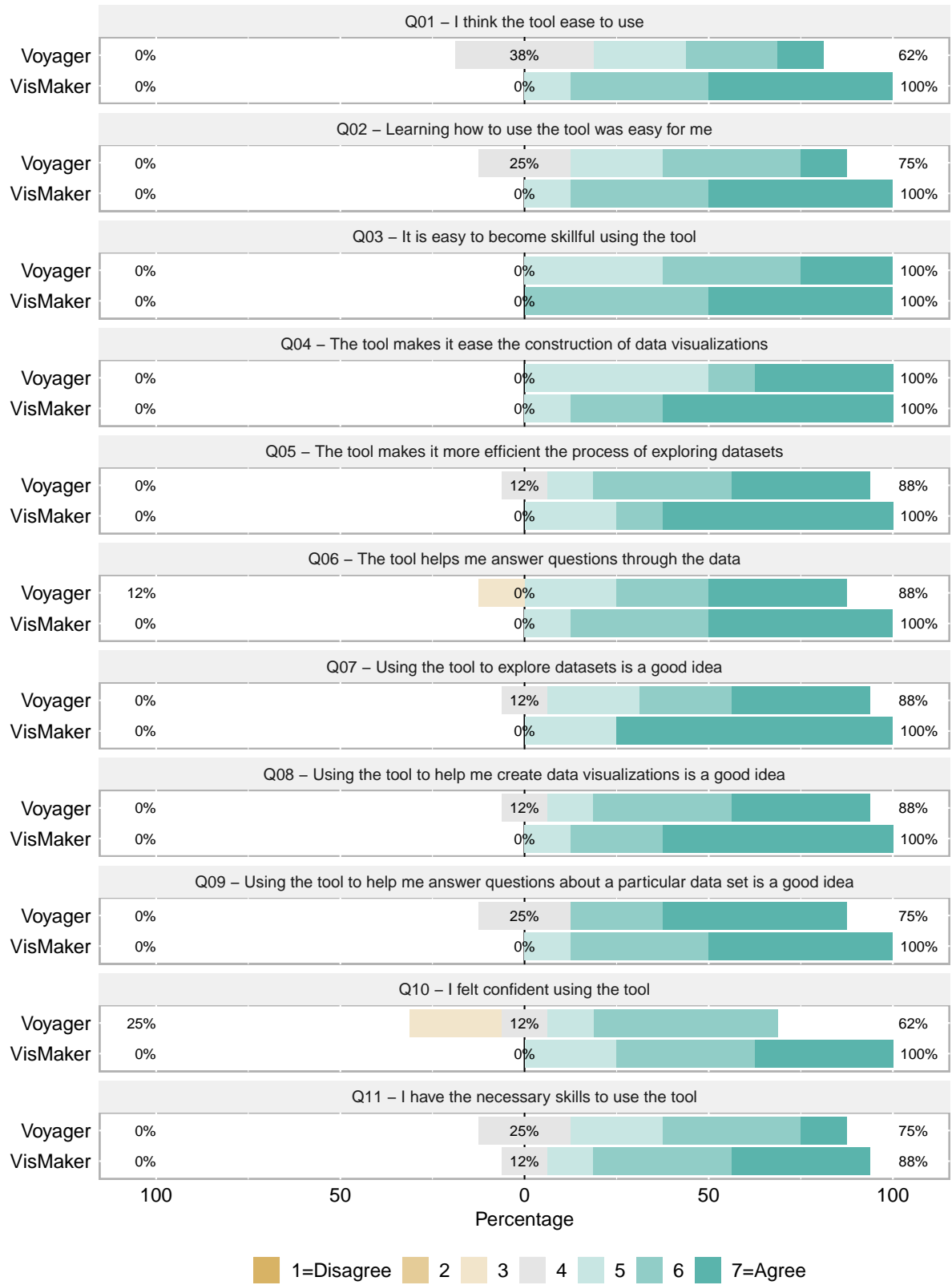}
	\caption{Questionnaire answers of study 2}
	\label{fig:questions_2}
\end{figure}

Similar to Study 1, we used the TAM questionnaire and applied the Mann-Whitney test. Although VisMaker obtained a higher median score in most questions, we only obtained statistical significance for Q01. Figure~\ref{fig:questions_2} and Table~\ref{tab:statistics-2} present these results.

\begin{table}[ht]
\centering
\caption{Results of Mann-Whitney statistical test for Study 2}
\label{tab:statistics-2}
\begin{tabular}{@{}lrrr@{}}
\toprule
\textbf{Question} & \textbf{VisMaker median} & \textbf{Voyager median} & \textbf{p-value} \\ \midrule
Q01 & 6.5 & 5 & \textbf{0.033*} \\
Q02 & 6.5 & 5.5 & 0.061 \\
Q03 & 6.5 & 6 & 0.126 \\
Q04 & 7 & 5.5 & 0.207 \\
Q05 & 7 & 6 & 0.460 \\
Q06 & 6.5 & 6 & 0.401 \\
Q07 & 7 & 6 & 0.219 \\
Q08 & 7 & 6 & 0.331 \\
Q09 & 6.5 & 6.5 & 0.775 \\
Q10 & 6 & 5.5 & 0.068 \\
Q11 & 6 & 5.5 & 0.250 \\ \bottomrule
\end{tabular}
\end{table}

\subsubsection{Participants' feedback}

We list below some feedback given by the participants for the two main questions we did in the interview of this study:

\textbf{\textit{Considering the recommendation approaches, which tool most helped you complete the task?}}
\begin{itemize}
    \item[P01:] \textit{``I found \textbf{VisMaker easier to view the charts}. The questions helped me.''}
    \item[P02:] \textit{``\textbf{VisMaker was much easier}, perhaps because it had fewer options and because I found the dataset cooler.''}
    \item[P03:] \textit{``\textbf{VisMaker was simpler}, visualizations were easier to understand, and it helped me more. Voyager had much information on the screen, and it just got a little confusing.''}
    \item[P04:] \textit{``I liked the recommendations because they gave new visualization ideas, and \textbf{I liked the way VisMaker presents the recommendations}. Voyager has too much information, and it confused me up a bit.''}
    \item[P05:] \textit{``I think \textbf{VisMaker}, but not necessarily because of the questions. In general, the \textbf{less information on the screen}, the better for me.''}
    \item[P06:] \textit{``\textbf{VisMaker helped me more, maybe because of the questions}. Sometimes I wanted to add more information, and I didn't know how so he would show the questions, and I knew better how to do it.''}
    \item[P07:] \textit{``\textbf{VisMaker was more interesting because it suggested hypotheses through questions}. It is friendlier and perhaps a better option for less experienced users.''}
    \item[P08:] \textit{``Watching the videos, I thought Voyager would be better but when using the tools \textbf{VisMaker was easier, maybe because I used VisMaker later}.''}
\end{itemize}
    
\noindent \textbf{\textit{Regarding the questions that VisMaker presented, you used them? Did they help you during the exploration process?}}
\begin{itemize}
    \item[P01:] \textit{``I found \textbf{VisMaker easier to view the charts}. I had the questions there, and they helped me understand things better. The questions helped me.''}
    \item[P02:] \textit{``I looked a few times at \textbf{the questions. They didn't make much difference to me}. I hardly used them.''}
    \item[P03:] \textit{``\textbf{I hardly used the questions}.''}
    \item[P04:] \textit{``\textbf{I think the questions are interesting} because they contextualize the recommendations, and I think it would help for laypeople and when the user doesn't know exactly what they are looking for.''}
    \item[P05:] \textit{``Sometimes I would read the \textbf{questions, and they would help me better understand the visualizations} that were being recommended and to think of new visualizations that I had not yet imagined.''}
    \item[P06:] \textit{``I read some questions. I looked at the \textbf{variables, and if I saw that they interested me, I read the whole question}.''}
    \item[P07:] \textit{``Yes, sometimes I read the questions. \textbf{They helped me to hypothesize about the data set}.''}
    \item[P08:] \textit{``\textbf{The questions helped a lot}. It was easier and faster to understand what was in the charts, and the questions helped me make questions and whether I really wanted to use the visualizations.''}
\end{itemize}

\section{Discussion}
\label{sec:discussion}

Study 1 showed that both tools enable and simplify the construction of statistical data visualizations. With a question in mind, participants quickly identified the variables they would need to use and mapped the necessary variables onto the different visualization channels.

In general, participants experienced some difficulty while using the tools on specific issues and made mistakes during the construction of the visualizations or even during the interpretation process of the constructed charts. However, despite these problems, most participants were able to complete most tasks using both tools. 

The answers to the questionnaires about the tools showed that VisMaker was significantly superior in terms of ease of use (Q01). Study 1 also reached statistical significance for questions Q02, Q03, Q07, and Q10, which address, respectively, the themes of ease, skill, use of the tool to explore datasets, and the confidence that the user had when using the tool.

Some of the feedback provided by the participants also attest to the ease that many of them felt when using VisMaker. But what factors led users to find VisMaker easier to use? What made Voyager less easy? The answer to these questions varies for each participant. In study 1, participant P01 stated that VisMaker's recommendations were more intuitive because they presented a question that helped them identify what they were looking for. Participant P05 stated that, in VisMaker, the questions made the recommendations more consistent. Participants P06, P07, P08, P09, P10, P11, and P15 also commented directly on the facility that VisMaker provided them through the questions. Other participants, such as P12, P14, and P16, commented on other aspects, such as the highlight by variable type in VisMaker, and the vast amount of information presented in Voyager, stating that it hindered them a little.

In Study 2, participants P02, P03, P04, and P05 commented on the smaller number of options that the VisMaker interface brought and on the vast amount of information that, in some cases, confused them when using Voyager. \textit{``In general, the less information on the screen, the better for me''}, said P05. Participants also commented about other positive aspects regarding the VisMaker recommendation approach. Participant P08, for example, stated that the questions helped them to understand the meaning of the recommended charts more easily. \textit{``VisMaker was more interesting because it suggested hypotheses through questions''}, also said participant P07 about VisMaker recommendations.

Although little used or even ignored by two participants (P02 and P03), the questions did not hinder them in carrying out their tasks. This was somewhat expected, as the participants did not have a specific question to answer.

\section{Concluding Remarks}
\label{sec:conclusion}

In this paper, we presented VisMaker, a visualization recommender tool that uses questions to facilitate the interpretation of the recommended visualizations. VisMaker comprises a visualization recommender system based on the use of rules that map, for the different combinations of types of variables, a set of visualizations considered more appropriate, in the sense of being related to the visualization the user has manually built. The main distinguishing factor of our tool is the use of questions generated based on the types of variables. The purpose underlying those questions is to facilitate and improve the understanding of the recommendations, the ability to raise hypotheses about the data, and the process of exploring the data.

We evaluated VisMaker through a comparison with the Voyager~2 tool to identify how the questions could help users to more easily obtain useful visualizations, assisting the data exploration process. The obtained results and feedback indicate that both VisMaker and Voyager~2 tool enabled users to accomplish the proposed tasks in both studies. They also indicate that the VisMaker recommendation approach can facilitate the understanding of the recommendations and help to make hypotheses about the data.

Some of our findings were inconclusive. Because our study involved few users, as future work we intend to carry out new studies with greater participation of users to verify the statistical significance of questions related to other aspects of the tools.

However, through the studies we conducted, we have already identified opportunities for future work, to improve the recommendation approach presented in VisMaker. The use of domain ontologies and machine learning models, such as those presented in some related works, can improve recommendations, helping to identify the efficiency of certain types of visualizations according to the characteristics of the data to be presented. The implementation of a recommender system that takes users' behavior into account can also improve recommendations, adapting the recommended visualizations to the taste of the users according to their preferences.

\bibliographystyle{unsrt}  
\bibliography{references}  

\end{document}